\documentclass[pra,amsmath,amssymb,showpacs,superscriptaddress,twocolumn]{revtex4}

\usepackage{graphicx}
\usepackage{dcolumn}
\usepackage[hypertex]{hyperref}
\usepackage{bm}
\usepackage[usenames,dvipsnames]{color}
\usepackage{amsmath}
\usepackage{amsfonts}
\usepackage{amsthm}
\usepackage{amssymb}
\usepackage{mathrsfs}
\usepackage{subfigure}
\usepackage{ulem}

\newtheorem{theorem}{Uncertainty equality}

\begin{document}

\title{Implications and applications of the variance-based uncertainty equalities}

\author{Yao Yao}
\email{yaoyao@csrc.ac.cn}
\affiliation{Beijing Computational Science Research Center, Beijing, 100084, China}
\affiliation{Synergetic Innovation Center of Quantum Information and Quantum Physics, University of Science and Technology of China,
Hefei, Anhui 230026, China}

\author{Xing Xiao}
\affiliation{Beijing Computational Science Research Center, Beijing, 100084, China}

\author{Xiaoguang Wang}
\affiliation{Zhejiang Institute of Modern Physics, Department of Physics, Zhejiang University, Hangzhou 310027, China}

\author{C. P. Sun}
\email{cpsun@csrc.ac.cn}
\affiliation{Beijing Computational Science Research Center, Beijing, 100084, China}
\affiliation{Synergetic Innovation Center of Quantum Information and Quantum Physics, University of Science and Technology of China,
Hefei, Anhui 230026, China}

\date{\today}

\begin{abstract}
In quantum mechanics, the variance-based Heisenberg-type uncertainty relations are a series of mathematical \textit{inequalities}
posing the fundamental limits on the achievable accuracy of the state preparations. In contrast, we construct and formulate two quantum uncertainty \textit{equalities}, which hold for all pairs of incompatible observables and indicate the new uncertainty relations recently introduced by L. Maccone and A. K. Pati
[Phys. Rev. Lett. \textbf{113}, 260401 (2014)]. Furthermore, we present an explicit interpretation lying behind the derivations and
relate these relations to the so-called \textit{intelligent states}. As an illustration, we investigate the properties of these uncertainty inequalities in the qubit system and a state-independent bound is obtained for the sum of variances. Finally, we apply these inequalities to the spin squeezing scenario
and its implication in interferometric sensitivity is also discussed.
\end{abstract}


\maketitle
\section{INTRODUCTION}
Similar to quantum entanglement, the uncertainty principle is also one of the characteristic traits of quantum mechanics and
is a fundamental departure form the principles of classical physics. Any pair of incompatible observables admit a certain
form of uncertainty relationship (e.g., an uncertainty inequality) and this constraint set ultimate bounds on the measurement precision
achievable for these quantities. Since Heisenberg introduced the first uncertainty relation about the product of the standard deviations
of canonical operators in 1927 \cite{Heisenberg1927,Kennard1927}, the scientific community has raised the long-standing controversy over
how to interpret and formulate the Heisenberg's original spirit \cite{Busch2007,Busch2014a,Wehner2010}.

Especially in recent heated debate, a series of novel error-tradeoff or measurement-disturbance relations have been proposed and
the community's enthusiasm on the uncertainty principle has been reactivated \cite{Ozawa2003,Ozawa2004,Hall2004,Busch2013,Weston2013,Branciard2013,
Busch2014b,Branciard2014,Buscemi2014,Bastos2014,Dressel2014,Lu2014}. However, the conventional variance-based uncertainty relations
possess a clear physical conception and still find a variety of applications in quantum information science, such as entanglement detection
\cite{Hofmann2003,Guhne2004}, quantum spin squeezing \cite{Walls1981,Wodkiewicz1985,Wineland1992,Kitagawa1993,Ma2011}, and even quantum metrology
\cite{Giovannetti2004,Giovannetti2006,Giovannetti2011}. In fact, it is precisely because of the uncertainty relations that quantum theory imposes
definite limits on the precision of measurement and the celebrated quantum Cram\'{e}r-Rao bound can also be deduced from the
Schr\"{o}dinger-Robertson uncertainty relation \cite{Hotta2004,Zhong2014}.

Intuitively, it is a well-accepted mathematical structure that the convectional uncertainty relations provide lower bounds to the product
or sum of the variances of incompatible Hermitian operators. Among the candidates, the most famous and popular form is the Robertson
uncertainty relation (RUR) \cite{Robertson1929}
\begin{equation}
\Delta A \Delta B\geq\left|\frac{1}{2i}\langle[A,B]\rangle\right|,\label{RUR}
\end{equation}
where the standard deviation $\Delta \mathcal{O}$ and expectation value $\langle\mathcal{O}\rangle$ are taken over the state $|\Psi\rangle$.
It is notable that the RUR can be derived from a slightly strengthened inequality, the Schr\"{o}dinger uncertainty relation (SUR) \cite{Schrodinger1930}
\begin{equation}
\Delta A^2 \Delta B^2\geq\left|\frac{1}{2i}\langle[A,B]\rangle\right|^2+\left|\frac{1}{2}\langle\{\breve{A},\breve{B}\}\rangle\right|^2,\label{SUR}
\end{equation}
where we define the operator $\breve{\mathcal{O}}=\mathcal{O}-\langle\mathcal{O}\rangle I$ and $I$ is the identity operator.

However, both the RUR and SUR suffer from the problem that they may be trivial even when $A$ and $B$ are incompatible on the state $|\Psi\rangle$,
for instance, $|\Psi\rangle$ is an eigenstate of either $A$ or $B$. In order to fix this flaw, recently Maccone and Pati presented two stronger
uncertainty relations based on the \textit{sum} of variances and these inequalities are guaranteed to be nontrivial whenever $|\Psi\rangle$
is not a common eigenstate of $A$ and $B$. The novel lower bound can be represented in a combination of both inequalities \cite{Maccone2014}
\begin{equation}
\Delta A^2+\Delta B^2\geq\max\{\mathcal{L}_1,\mathcal{L}_2\},
\end{equation}
where we define
\begin{align}
\mathcal{L}_1&=\pm i\langle[A,B]\rangle+|\langle\Psi|A\pm iB|\Psi^\perp\rangle|^2,\label{inequality1}\\
\mathcal{L}_2&=\frac{1}{2}|\langle\Psi^\perp_{A+B}|A+B|\Psi\rangle|^2,\label{inequality2}
\end{align}
Here $|\Psi^\perp\rangle$ is an arbitrary state orthogonal to $|\Psi\rangle$ and $|\Psi^\perp_{A+B}\rangle$ is specified according to
the Vaidman's formula \cite{Vaidman1992,Goldenberg1996}
\begin{equation}
\mathcal{O}|\Psi\rangle=\langle\mathcal{O}\rangle|\Psi\rangle+\Delta\mathcal{O}|\Psi^\perp_{\mathcal{O}}\rangle.
\end{equation}
Moreover, utilizing the same techniques employed to derive (\ref{inequality1}), Maccone and Pati also obtained an amended RUR \cite{Maccone2014}
\begin{equation}
\Delta A \Delta B\geq\pm \frac{i}{2}\langle[A,B]\rangle/\left(1-\frac{1}{2}\left|\langle\Psi|\frac{A}{\Delta A}\pm i\frac{B}{\Delta B}|\Psi^\perp\rangle\right|^2\right).\label{inequality3}
\end{equation}

In this work, we try to look at such a problem from another perspective. Given two noncommuting operators $A$ and $B$, we can define
the uncertainty functional $\mathcal{U}(\Psi)=\Delta A^2\Delta B^2$. Indeed, the RUR and SUR follow directly from the \textit{uncertainty equality}
\begin{equation}
\mathcal{U}(\Psi)=\left|\frac{1}{2i}\langle[A,B]\rangle\right|^2+\left|\frac{1}{2}\langle\{\breve{A},\breve{B}\}\rangle\right|^2+\mathcal{R}(\Psi),
\end{equation}
where $\mathcal{R}(\Psi)$ is a positive semidefinite remainder term, emerging from the application of the Cauchy-Schwarz inequality to
$\langle \breve{A}^2\rangle\langle \breve{B}^2\rangle$. Can we construct other uncertainty equalities for $\mathcal{U}(\Psi)$ and another functional
$\mathcal{W}(\Psi)=\Delta A^2+\Delta B^2$, which can straightforward lead to the inequalities derived in Ref. \cite{Maccone2014} ?
Here we show that the answer is affirmative and elucidate the physical meaning behind these inequalities.

An outline of the reminder of the paper is as follows. In Sec. \ref{sec2}, we construct and formulate two quantum uncertainty \textit{equalities},
which hold for all pairs of incompatible observables and imply the new uncertainty relations introduced by Maccone and Pati.
Furthermore, we present an explicit interpretation lying behind the derivations and relate these relations to the so-called intelligent states.
In Sec. \ref{sec3}, we investigate the properties of these uncertainty inequalities in the qubit system and
a state-independent bound is obtained for the sum of variances. In Sec. \ref{sec4}, we apply these inequalities to the spin squeezing scenario
and its implication in interferometric sensitivity is also discussed. Finally, Sec. \ref{sec5} is devoted to the discussion and conclusion.

\section{Uncertainty equalities imply uncertainty relations}\label{sec2}
\subsection{New uncertainty equalities}\label{sec2a}
As indicated in Ref. \cite{Maccone2014}, the lower bound $\mathcal{L}_2$ is derived from the uncertainty equality
\begin{equation}
\Delta A^2+\Delta B^2=\frac{1}{2}\left[\Delta(A+B)^2+\Delta(A-B)^2\right].\label{equality3}
\end{equation}
In fact, we can obtain another lower bound
\begin{equation}
\mathcal{L}_3=\frac{1}{2}\Delta(A-B)^2=\frac{1}{2}|\langle\Psi^\perp_{A-B}|A-B|\Psi\rangle|^2.
\end{equation}
In the following, we first construct and prove two uncertainty equalities which imply the uncertainty inequalities (\ref{inequality1}) and (\ref{inequality3}).
Note that here we refer the lower bounds as to the corresponding uncertainty relations.

\begin{theorem}
\begin{equation}
\mathcal{W}(\Psi)=\pm i\langle[A,B]\rangle+\sum_{k=1}^{d-1}\left|\langle\Psi|A\pm iB|\Psi^\perp_{k}\rangle\right|^2,\label{equality1}
\end{equation}
where $\mathcal{W}(\Psi)=\Delta A^2+\Delta B^2$ and $\{|\Psi\rangle,|\Psi^\perp_{k}\rangle_{k=1}^{d-1}\}$
comprise an orthonormal complete basis in the $d$-dimensional Hilbert space.
\end{theorem}

\textit{Proof.} For simplicity, let us define the operator $\Pi=I-|\Psi\rangle\langle\Psi|$ and the state $|\chi^{\pm}\rangle=(A\pm iB)|\Psi\rangle$.
Note that $\Pi^2=\Pi$, which is a projector of the L\"{u}ders type \cite{Luders2006}. The $\pm$ sign in $|\chi^{\pm}\rangle$ is due to the symmetry
between $A$ and $B$ since $\mathcal{W}(\Psi)$ must be invariant under $A\Leftrightarrow B$ (see below). We have
\begin{align}
\langle\chi^{\mp}|\Pi|\chi^{\mp}\rangle&=\langle\Psi|(A\pm iB)(I-|\Psi\rangle\langle\Psi|)(A\mp iB)|\Psi\rangle \nonumber\\
&=\langle\chi^{\mp}|\chi^{\mp}\rangle-\langle\chi^{\mp}|\Psi\rangle\langle\Psi|\chi^{\mp}\rangle \nonumber\\
&=\langle A^2+B^2\mp i[A,B]\rangle  \nonumber\\
&\qquad -(\langle A\rangle\pm i\langle B\rangle)(\langle A\rangle\mp i\langle B\rangle) \nonumber\\
&=\Delta A^2+\Delta B^2\mp i\langle[A,B]\rangle.\label{ancilla1}
\end{align}
Since $\Pi$ is the orthogonal complement to $|\Psi\rangle\langle\Psi|$ (e.g, $\langle\Psi|\Pi|\Psi\rangle$=0),
we can choose an \textit{arbitrary} orthogonal decomposition of the projector $\Pi$
\begin{equation}
\Pi=\sum_{k=1}^{d-1}|\Psi^{\perp}_k\rangle\langle\Psi^{\perp}_k|,\label{ancilla2}
\end{equation}
where $\{|\Psi\rangle,|\Psi^\perp_{k}\rangle_{k=1}^{d-1}\}$ comprise an orthonormal complete basis in the $d$-dimensional Hilbert space.
Combining Eqs. (\ref{ancilla1}) and (\ref{ancilla2}), we obtain the uncertainty relation (\ref{equality1}). \hfill $\blacksquare$

\begin{theorem}
\begin{equation}
\mathcal{U}(\Psi)^{1/2}=\frac{\pm \frac{i}{2}\langle[A,B]\rangle}{1-\frac{1}{2}\sum_{k=1}^{d-1}\left|\langle\Psi|\frac{A}{\Delta A}\pm i\frac{B}{\Delta B}|\Psi^\perp_k\rangle\right|^2},
\label{equality2}
\end{equation}
where $\mathcal{U}(\Psi)=\Delta A^2\Delta B^2$ and $\{|\Psi\rangle,|\Psi^\perp_{k}\rangle_{k=1}^{d-1}\}$
comprise an orthonormal complete basis in the $d$-dimensional Hilbert space.
\end{theorem}

\textit{Proof.}  Similar to the above arguments, first define the unnormalized state vector $|\xi^{\pm}\rangle=(\frac{A}{\Delta A}\pm i\frac{B}{\Delta B})|\Psi\rangle$.
We have the identity
\begin{align}
\langle\xi^{\mp}|\Pi|\xi^{\mp}\rangle&=\langle\xi^{\mp}|\xi^{\mp}\rangle-\langle\xi^{\mp}|\Psi\rangle\langle\Psi|\xi^{\mp}\rangle \nonumber\\
&=\left\langle \frac{A^2}{\Delta A^2}+\frac{B^2}{\Delta B^2}\mp \frac{i[A,B]}{\Delta A\Delta B}\right\rangle  \nonumber\\
&\qquad -\left(\frac{\langle A\rangle}{\Delta A}\pm i\frac{\langle B\rangle}{\Delta B}\right)\left(\frac{\langle A\rangle}{\Delta A}\mp i\frac{\langle B\rangle}{\Delta B}\right) \nonumber\\
&=2\mp i\frac{\langle[A,B]\rangle}{\Delta A\Delta B}.\label{ancilla3}
\end{align}
From Eqs. (\ref{ancilla2}) and (\ref{ancilla3}), we obtain the uncertainty relation (\ref{equality2}).
Note that we always assume that $\Delta A\Delta B\neq0$, e.g., $|\Psi\rangle$ is not an eigenstate of either $A$ or $B$. \hfill $\blacksquare$

Before proceeding, some remarks can be made on the significance of the above two uncertainty equalities.
First, if we retain only one term associated with $|\Psi^{\perp}\rangle\in\{|\Psi^\perp_{k}\rangle_{k=1}^{d-1}\}$ in the summation and
discard the others, the uncertainty equalities (\ref{equality1}) and (\ref{equality2}) reduce to the uncertainty inequalities (\ref{inequality1}) and (\ref{inequality3}), respectively.
It is worth emphasizing that in contrast to the derivations in Ref. \cite{Maccone2014}, the Cauchy-Schwarz inequality is not
involved here. Moreover, the tightness of the inequality is indicated by the uncertainty equality. For example, when
$|\Psi^{\perp}\rangle$ is of the form ($\mathcal{N}$ is the normalization factor)
\begin{align}
|\Psi^{\perp}\rangle=(A\mp iB-\langle A\mp iB\rangle)|\Psi\rangle/\mathcal{N}
=\Pi|\chi^{\mp}\rangle/\mathcal{N},
\end{align}
it is easy to see that the contribution from all other terms in the summation of (\ref{equality1}) vanishes
\begin{align}
\sum_{k=2}^{d-1}\left|\langle\Psi|A\pm iB|\Psi^\perp_{k}\rangle\right|^2=\mathcal{N}\sum_{k=2}^{d-1}\left|\langle\Psi^{\perp}_1|\Psi^\perp_{k}\rangle\right|^2=0,
\end{align}
where we assume $|\Psi^{\perp}\rangle=|\Psi^{\perp}_1\rangle$. This result coincides with that of \cite{Maccone2014}.
Therefore, in view of Ref. \cite{Maccone2014}, we can naturally interpret the tightness of (\ref{inequality1}) or (\ref{inequality3}) as imposing
constraints on the properties of $|\Psi^{\perp}\rangle$.

\subsection{Interpretations of uncertainty inequalities}
However, we still wonder why these seemingly curious expressions, such as $A\pm iB$ and $\frac{A}{\Delta A}\pm i\frac{B}{\Delta B}$,
appear in these inequalities. Here we provide an explicit interpretation of (\ref{inequality1}) and (\ref{inequality3}), and
the critical role of the \textit{intelligent states} is highlighted. From this perspective, two significant types of states
should be introduced: the ordinary intelligent states (OISs) provide an equality in the RUR \cite{Aragone1974,Aragone1976},
while the generalized intelligent states (GISs) do the same in the SUR \cite{Trifonov1994}.
It is clear that the OISs form a subset of the GISs and the set of OISs is unitarily equivalent to the set of GISs \cite{Nha2007}.
Most importantly, owing to the application of the Cauchy-Schwarz inequality, the GISs for operators $A$ and $B$
must satisfy the following characteristic eigenvalue equation \cite{Trifonov1994,Puri1994}
\begin{equation}
(A+i\gamma B)|\Psi\rangle=\lambda|\Psi\rangle,\label{condition}
\end{equation}
where $\gamma$ is an arbitrary \textit{complex number} (e.g., $\gamma\in\mathbb{C}$) and the eigenvalue $\lambda=\langle A\rangle+i\gamma\langle B\rangle$.
For the particular case of real $\gamma\in\mathbb{R}$, the eigenvalue equation (\ref{condition}) determines the OISs for operators $A$ and $B$ \cite{Jackiw1968}.
In addition, it should be emphasized that the concept of OISs is not equivalent to \textit{minimum-uncertainty states} (MUSs) in general
\cite{Wodkiewicz1985,Aragone1974,Aragone1976,Jackiw1968} and we also obtain a constraint for $|\Psi\rangle$ which should be satisfied
if $\mathcal{W}(\Psi)$ is to be a minimum
\begin{equation}
(\breve{A}^2+\breve{B}^2)|\Psi\rangle=(\Delta A^2+\Delta B^2)|\Psi\rangle.
\end{equation}
For more details, see Appendix \ref{appendix1}.

For further discussion, the necessary condition (\ref{condition}) can be rewritten as
\begin{equation}
(\breve{A}+i\gamma \breve{B})|\Psi\rangle=0.\label{condition1}
\end{equation}
By multiplying $\breve{A}+i\gamma \breve{B}$ or $\breve{A}-i\gamma \breve{B}$ upon (\ref{condition1}), we have the following two equations \cite{Trifonov1994,Puri1994}
\begin{align}
\Delta A^2-\gamma^2\Delta B^2&=-i\gamma\langle F\rangle,\label{condition2}\\
\Delta A^2+\gamma^2\Delta B^2&=\gamma\langle C\rangle,\label{condition3}
\end{align}
where we define the Hermitian operators
\begin{equation}
C=-i[A,B]=-i[\breve{A},\breve{B}],\quad F=\{\breve{A},\breve{B}\}.
\end{equation}
Therefore, the solution to (\ref{condition2}) and (\ref{condition3}) is
\begin{equation}
\gamma=\frac{\langle C\rangle+i\langle F\rangle}{2\Delta B^2},\quad |\gamma|^2=\frac{\Delta A^2}{\Delta B^2},
\end{equation}
where we still ignore the trivial cases and assume $\Delta A\Delta B\neq0$.

Since the uncertainty relations (\ref{inequality1}) and (\ref{inequality3}) are both extensions to the RUR,
we should concentrate on the special case of $\gamma\in\mathbb{R}$, that is, $\gamma=\pm\Delta A / \Delta B$. Therefore, the eigenvalue equation
(\ref{condition}) can be recast as
\begin{equation}
\left(\frac{A}{\Delta A}\pm i\frac{B}{\Delta B}\right)|\Psi\rangle=\frac{\lambda}{\Delta A}|\Psi\rangle.\label{condition4}
\end{equation}
Thus, if we define the following two quantities
\begin{align}
\Theta_1&=\left|\langle\Psi|\frac{A}{\Delta A}\pm i\frac{B}{\Delta B}|\Psi^\perp\rangle\right|^2,\\
\Theta_2&=\left|\langle\Psi|A\pm iB|\Psi^\perp\rangle\right|^2,
\end{align}
it soon becomes clear that the value of $\Theta_1$ reveals \textit{the extent to which} $|\Psi\rangle$ deviates from being an OIS, or more precisely,
$\Theta_1$ characterizes \textit{the extent to which} $\Delta A \Delta B$ deviates from $|\langle[A,B]\rangle|/2$.
Meanwhile, by noticing the inequality
\begin{equation}
\Delta A^2+\Delta B^2\geq2\Delta A\Delta B\geq|\langle[A,B]\rangle|,
\end{equation}
we realize that if we require $\Delta A^2+\Delta B^2=|\langle[A,B]\rangle|$, the condition $\Delta A=\Delta B$ must be satisfied.
In this circumstance, the eigenvalue equation (\ref{condition4}) reduces to
\begin{equation}
( A\pm i B)|\Psi\rangle=\lambda|\Psi\rangle.
\end{equation}
Thus, we recognize that $\Theta_2$ characterizes \textit{the extent to which} $\Delta A^2+\Delta B^2$ deviates from $|\langle[A,B]\rangle|$.
It is worth noting that the above explanation does not depend on extra properties of $|\Psi^{\perp}\rangle$ except for $\langle\Psi|\Psi^{\perp}\rangle=0$,
which in turn leads to the uncertainty inequalities (\ref{inequality1}) and (\ref{inequality3}).

\section{Qubit system as an illustration}\label{sec3}
As the most commonly used building blocks for quantum information processing, qubit systems have played an irreplaceable role
not only in theoretical analysis but also in experimental tests due to its unique properties. Therefore, it would be of great interest
to evaluate the performance of these new uncertainty relations and to compare them with RUR or SUR in the context of qubit systems.

In the Bloch sphere representation, $|\Psi^{\perp}\rangle$ is \textit{unique} (up to an irrelevant overall phase factor) and
its Bloch vector is antiparallel with respect to that of $|\Psi\rangle$, which implies $\Pi=I-|\Psi\rangle\langle\Psi|=|\Psi^{\perp}\rangle\langle\Psi^{\perp}|$.
From the derivations in Sec. \ref{sec2a}, it turns out that the uncertainty inequalities (\ref{inequality1}) and (\ref{inequality3})
\textit{automatically} become equalities for arbitrary single-qubit pure states. In fact, we have
\begin{align}
\mathcal{U}(\Psi)^{1/2}&=\pm \frac{i}{2}\langle[A,B]\rangle/\left(1-\frac{1}{2}\Theta_1\right),\label{identity1}\\
\mathcal{W}(\Psi)&=\pm i\langle[A,B]\rangle+\Theta_2.\label{identity2}
\end{align}
In other words, the above identities indicate that the novel uncertainty relations (\ref{inequality1}) and (\ref{inequality3}) accounts for all the uncertainty
predicted by the sum and product of the standard deviations in qubit system.

In full generality, we consider two arbitrary Hermitian operators
\begin{align}
A=&\alpha_1 I+\alpha_2 \ \vec{a}\cdot\vec{\sigma},\\
B=&\beta_1 I+\beta_2 \ \vec{b}\cdot\vec{\sigma},
\end{align}
where $\{\alpha_i,\beta_i\}$ are real parameters, $\vec{a},\vec{b}\in\mathbb{R}^3$ are unit vectors and $\vec{\sigma}=(\sigma_x,\sigma_y,\sigma_z)$
are standard Pauli matrices. Meanwhile, the most general pure state of a single qubit is of the form (up to an unobservable phase factor)
\begin{equation}
|\Psi\rangle=\cos\frac{\theta}{2}|0\rangle+e^{i\varphi}\sin\frac{\theta}{2}|1\rangle,
\end{equation}
with $\theta\in[0,\pi]$ and $\varphi\in[0,2\pi]$. In Bloch representation, the corresponding density operator can be written as $\rho=|\Psi\rangle\langle\Psi|=\frac{1}{2}(I+\vec{r}\cdot\vec{\sigma})$
with the Bloch vector $\vec{r}=(\sin\theta\cos\varphi,\sin\theta\sin\varphi,\cos\theta)$. To simplify the problem, one may consider $\vec{a}\cdot\vec{\sigma}$ ($\vec{b}\cdot\vec{\sigma}$) instead of
$A$ ($B$) and this strategy is usually employed in the discussion of entropic uncertainty relations \cite{Ghirardi2003,Bosyk2012}. It is reasonable since $\vec{a}\cdot\vec{\sigma}$ and $A$
have the same eigenstates and the eigenvalues are not involved in the corresponding entropy functions. However, the standard deviation $\Delta A$ does depend on the eigenvalues.

Fortunately, this simplification still works here. In our notation, the RUR (\ref{RUR}) is represented as
\begin{equation}
|\alpha_2|\sqrt{1-(\vec{a}\cdot\vec{r})^2}|\beta_2|\sqrt{1-(\vec{b}\cdot\vec{r})^2}\geq|\alpha_2\beta_2||(\vec{a}\times\vec{b})\cdot\vec{r}|,
\end{equation}
which is equivalent to
\begin{equation}
\sqrt{1-(\vec{a}\cdot\vec{r})^2}\sqrt{1-(\vec{b}\cdot\vec{r})^2}\geq|(\vec{a}\times\vec{b})\cdot\vec{r}|.
\end{equation}
Thus, we can restrict our attention to the class of Hermitian operators of the forms $A=\vec{a}\cdot\vec{\sigma}$ and $B=\vec{b}\cdot\vec{\sigma}$
since the uncertainty inequalities (\ref{inequality1}) and (\ref{inequality3}) are both extensions of RUR \cite{note}. To Further simplify the discussion,
we can assume that $A$ and $B$ lie in the $x-y$ plane with loss of generality, that is
\begin{align}
A=&\cos\phi\ \sigma_x+\sin\phi\ \sigma_y,\\
B=&\sin\phi\ \sigma_x+\cos\phi\ \sigma_y,
\end{align}
where the angle between $A$ and $B$ is entirely characterized by the parameter $\phi$.

First, it is easy to verify that the identities (\ref{identity1}) and (\ref{identity2}) indeed hold for arbitrary pure states $|\Psi\rangle$
by utilizing $|\Psi^{\perp}\rangle\langle\Psi^{\perp}|=\frac{1}{2}(I-\vec{r}\cdot\vec{\sigma})$. In particular, we have
\begin{align}
\mathcal{W}(\Psi)&=\pm i\langle[A,B]\rangle+|\langle\Psi|A\pm iB|\Psi^\perp\rangle|^2 \nonumber\\
&=1+\cos^2\theta-\sin^2\theta\sin2\varphi\sin2\phi\nonumber\\
&\geq 1-|\sin2\phi|\nonumber\\
&=1-|\vec{a}\cdot\vec{b}|.
\end{align}
Alternatively, we can prove in Bloch formulism that the above inequality indeed provides a state-independent lower bound for the quadratic functional $\mathcal{W}(\Psi)=\Delta A^2+\Delta B^2$
by using the parallelogram law (see Appendix \ref{appendix2})
\begin{equation}
\mathcal{W}(\Psi)\geq1-|\vec{a}\cdot\vec{b}|=2(1-c^2),
\end{equation}
where $c=\max_{i,j}|\langle a_i|b_j\rangle|$ and $\{|a_i\rangle\}$ ($\{|b_j\rangle\}$) are the corresponding eigenvectors of $A$ ($B$).
Note that $c$ is the most common and important quantity in the formulation of entropic uncertainty relations \cite{Wehner2010}.

Since the inequalities (\ref{inequality1}) and (\ref{inequality3}) are saturated for any single-qubit pure state, then we focus on the
performance of the inequality (\ref{inequality2}) comparing with the RUR or SUR. We notice that reformulation by normalization turns out to be a
relatively reasonable way to compare different types of uncertainty relations, e.g., dividing both sides of the inequalities by their own lower bound
\cite{Kitagawa1986,Fujikawa2013,Baek2014}. According to this line of thought, we can define the following two functionals
\begin{align}
\mathcal{U}_1(\theta,\varphi,\phi)=&\frac{\Delta A^2 \Delta B^2}{|\langle[A,B]\rangle/2|^2},\\
\mathcal{U}_2(\theta,\varphi,\phi)=&\frac{\Delta A^2+\Delta B^2}{|\langle\Psi^\perp_{A+B}|A+B|\Psi\rangle|^2/2}.
\end{align}
Therefore, the performances (or tightnesses) of uncertainty relations is to compare the left hand sides of the inequalities
with the uniformly normalized lower bound $1$. In our notation, we have
\begin{align}
\mathcal{U}_1=&\frac{[1-\sin^2\theta\cos^2(\varphi-\phi)][1-\sin^2\theta\sin^2(\varphi+\phi)]}{\cos^2\theta\cos^22\phi},\label{ancilla4}\\
\mathcal{U}_2=&\frac{2-\sin^2\theta(1+\sin2\varphi\sin2\phi)}{(1+\sin2\phi)[1-\sin^2\theta(1+\sin2\varphi)/2]}.\label{ancilla5}
\end{align}

\begin{figure}[htbp]
\begin{center}
\includegraphics[width=0.40\textwidth ]{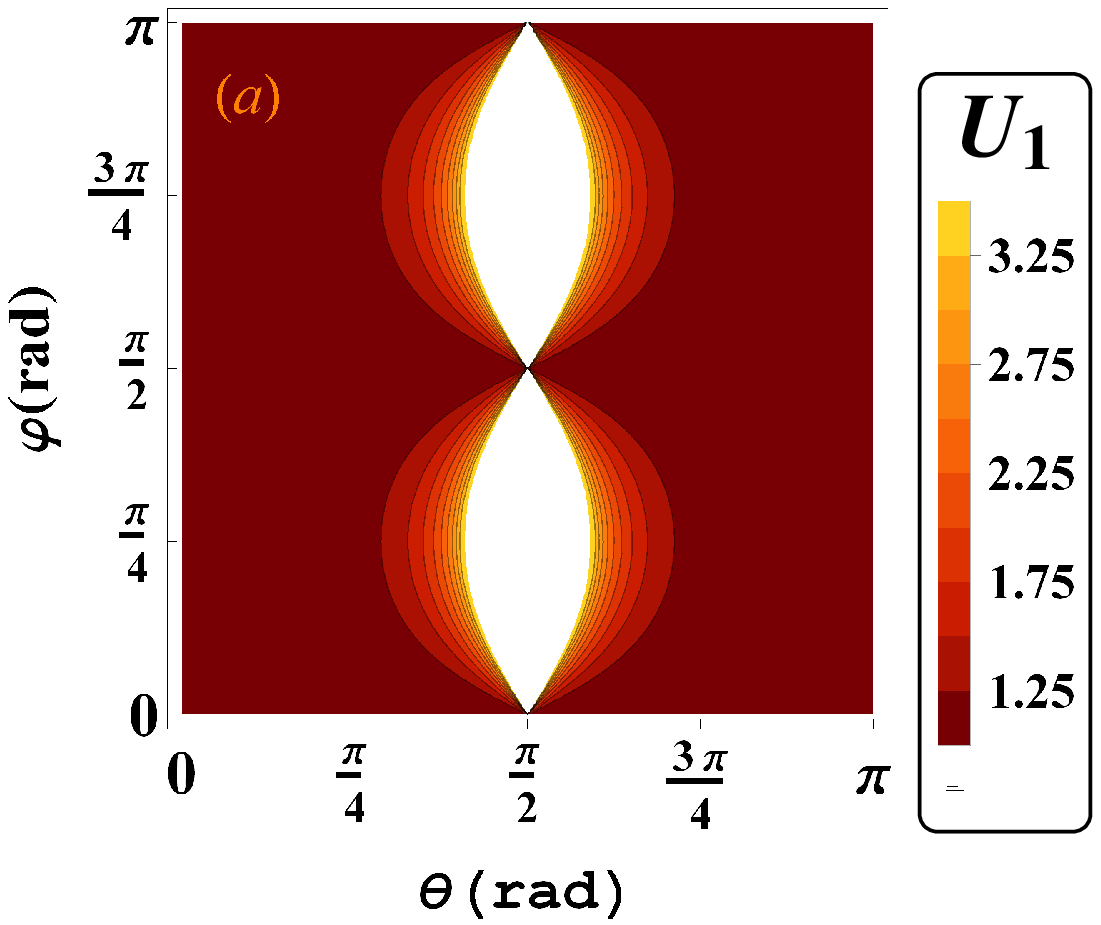}\\
\includegraphics[width=0.40\textwidth ]{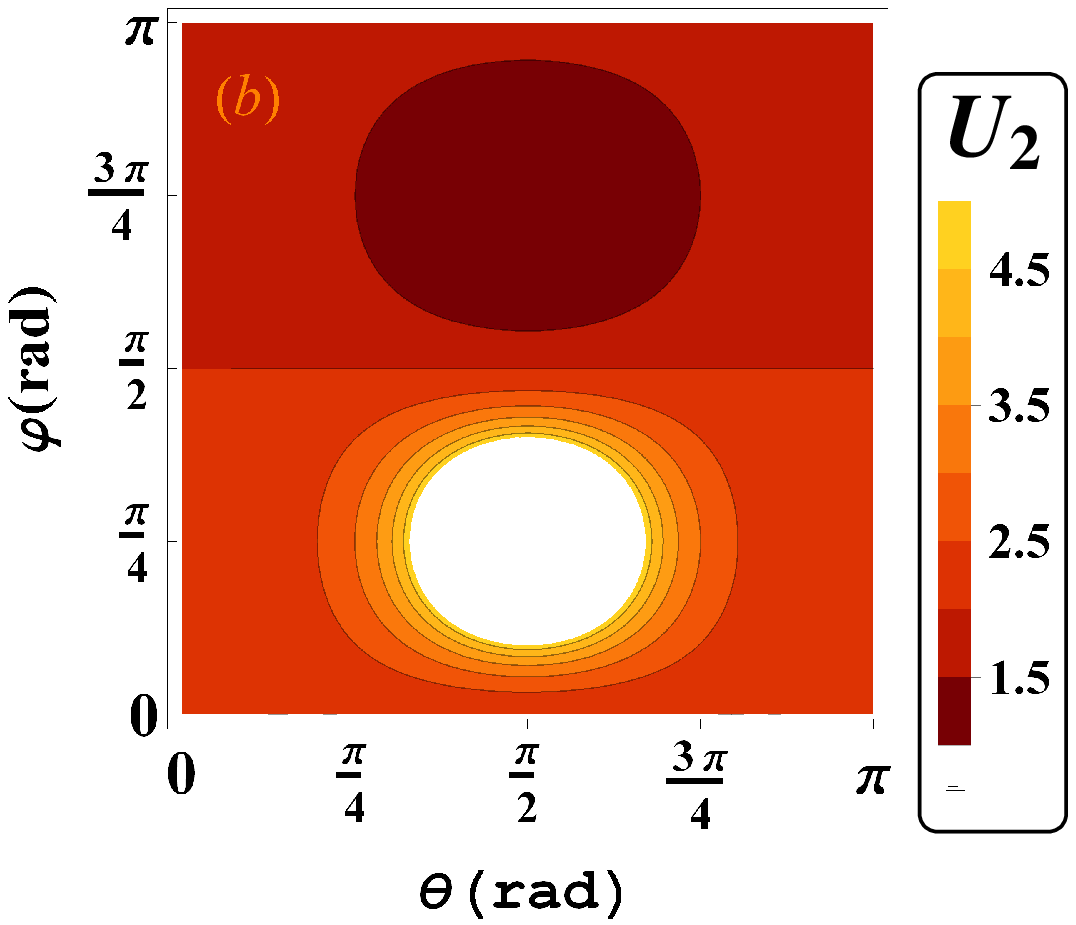}
\end{center}
\caption{(Color online) (a) The contour plot of $\mathcal{U}_1(\phi=0)$ as a function of polar angle $\theta$ and
azimuthal angle $\varphi$; (b) The contour plot of $\mathcal{U}_2(\phi=0)$ as a function of the parameters $\theta$ and $\varphi$.
Note that lighter regions show higher values of the functions.
}\label{comparison1}
\end{figure}

We first focus on the case $\phi=0$ where two observables $A=\sigma_x$ and $B=\sigma_y$ are complementary to each other.
Recall that the associated eigenvectors of the Pauli matrices are mutually unbiased bases of $\mathbb{C}^2$. In Fig. \ref{comparison1},
we show the contour plots of $\mathcal{U}_1(\phi=0)$ and $\mathcal{U}_2(\phi=0)$ as a function of polar angle $\theta$ and
azimuthal angle $\varphi$. We can easily check that an equality $\mathcal{U}_1=1$ holds for $\theta=0,\pi$ and
arbitrary $\varphi$ or arbitrary $\theta$ and $\varphi=n\pi/2$ ($n$ is an integer and notice the symmetry of the function $\mathcal{U}_1$).
When $\theta\rightarrow\pi/2$, $\mathcal{U}_1$ diverges since the denominator of $\mathcal{U}_1$ approaches $0$ near this critical region.
Meanwhile, $\mathcal{U}_2=1$ is fulfilled for $\theta=\pi/2$ and $\varphi=3\pi/4$ since $|\Psi\rangle=\frac{1}{\sqrt{2}}(|0\rangle+e^{3\pi i/4}|1\rangle)$
is one of the eigenvectors of $A-B=\sigma_x-\sigma_y$, while $\mathcal{U}_2$ diverges for $\theta=\pi/2$ and $\varphi=\pi/4$ due to
the fact that $\Delta\mathcal{O}$ vanishes if and only if $|\Psi\rangle$ is an eigenstate of the observable $\mathcal{O}$
(see the identity (\ref{equality3})).

\begin{figure}[htbp]
\begin{center}
\includegraphics[width=0.40\textwidth ]{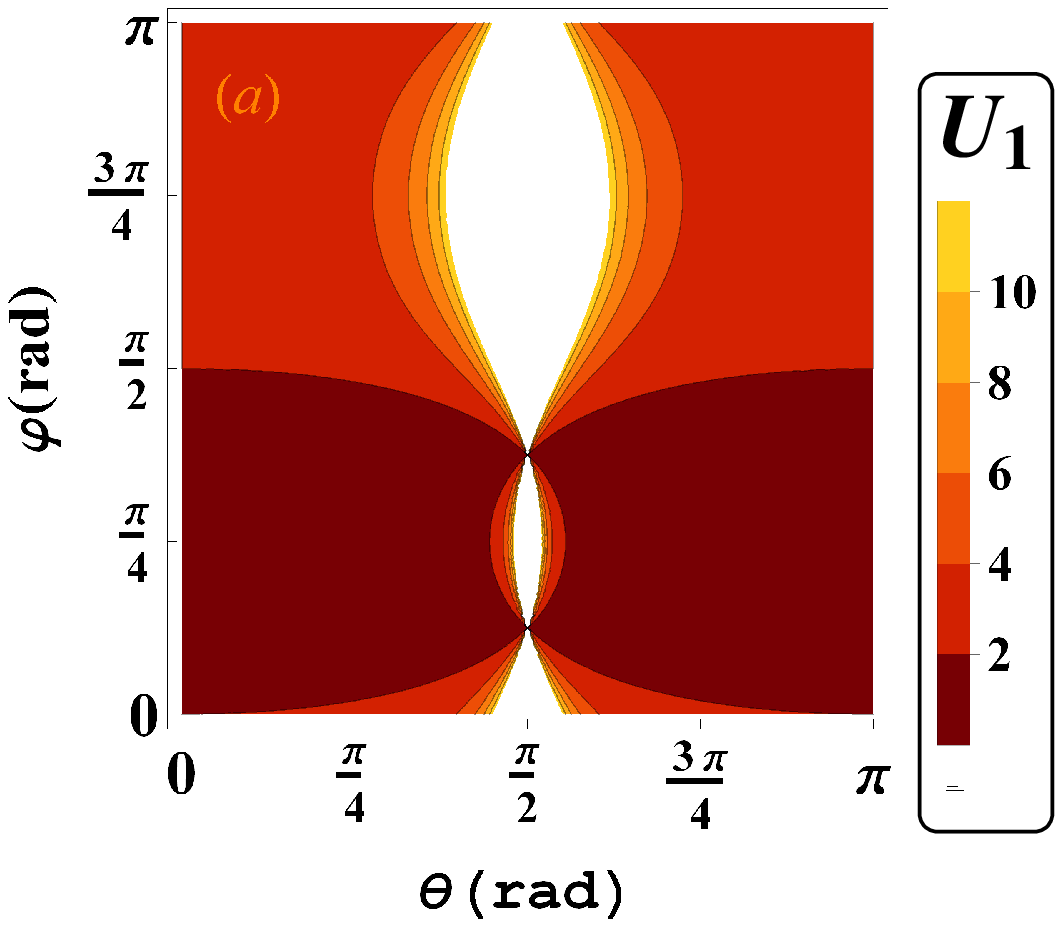}\\
\includegraphics[width=0.40\textwidth ]{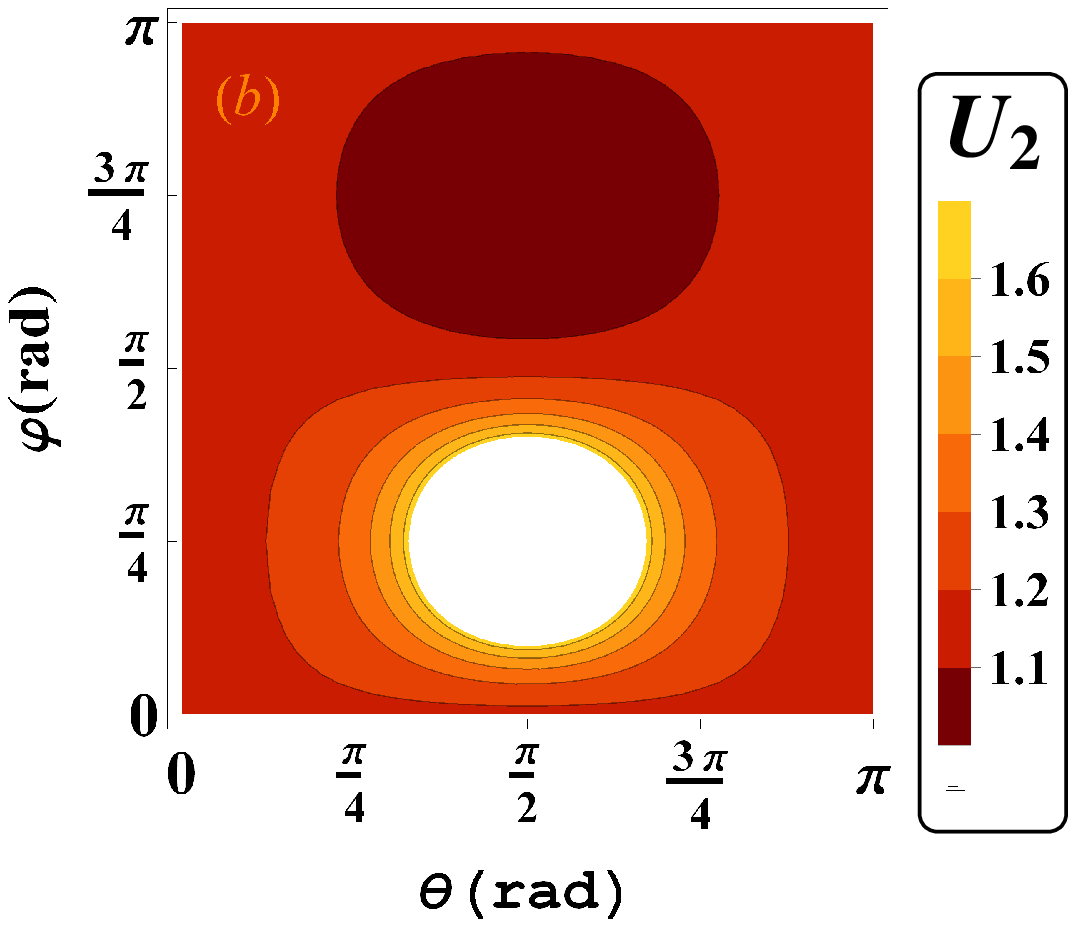}
\end{center}
\caption{(Color online) (a) The contour plot of $\mathcal{U}_1(\phi=\pi/8)$ as a function of polar angle $\theta$ and
azimuthal angle $\varphi$; (b) The contour plot of $\mathcal{U}_2(\phi=\pi/8)$ as a function of the parameters $\theta$ and $\varphi$.
Note that lighter regions show higher values of the functions.
}\label{comparison2}
\end{figure}

For comparison, we also present the contour plots for the case $\phi=\pi/8$. It is evident that the structure of $\mathcal{U}_1(\phi=\pi/8)$
is greatly different from that of $\mathcal{U}_1(\phi=0)$, but on the contrary $\mathcal{U}_2$ almost remains unchanged. Note that generally
$A-B=(\cos\phi-\sin\phi)(\sigma_x-\sigma_y)$, so that the condition for convergence or divergence of $\mathcal{U}_2$ is independent of
the value of $\phi$. Indeed, we can also interpret this result intuitively by noting that $\phi$ only appears in an overall multiplicative
factor of the denominator of $\mathcal{U}_2$ (see Eq. (\ref{ancilla5})). Therefore, in order to neatly avoid the divergence of $\mathcal{U}_2$, we can
reformulate the lower bound $\mathcal{L}_2$ as
\begin{equation}
\mathcal{L}'_2=\max\{\mathcal{L}_2,\mathcal{L}_3\}.
\end{equation}
For the sake of completeness, we additionally have plotted the corresponding uncertainty function $\mathcal{U}_3$ for the SUR and it
turns out that $\mathcal{U}_3$ never varies and is identically equal to unity, which means that the equality of the SUR always holds for arbitrary pure
states ($|\vec{r}|=1$). In fact, we have the following identity
\begin{align}
\Upsilon(A,B,\rho)=\left[1-(\vec{a}\cdot\vec{b})^2\right]\left(1-\vec{r}^2\right)\geq0,
\end{align}
where we define
\begin{equation}
\Upsilon(A,B,\rho)=\Delta A^2 \Delta B^2-\frac{1}{4}|\langle C\rangle|^2-\frac{1}{4}|\langle F\rangle|^2.
\end{equation}
Hence, the SUR can be employed in the domain of discrete variables to detect the mixedness of qubit states \cite{Mal2013}.

\section{Squeezed states and quantum metrology}\label{sec4}
As indicated in previous literature, the definition of squeezing or reduction of quantum fluctuations is intimately intertwined with
uncertainty relations \cite{Ma2011}. For instance, if two arbitrary observables $A$ and $B$ obey the commutation relation
$[A, B]=iC$ and the RUR, a state $|\Psi\rangle$ is said to be squeezed in $\mathcal{O}\in\{A,B\}$ if the uncertainty
in $\mathcal{O}$ satisfies the relation \cite{Walls1981,Wodkiewicz1985,Trifonov1994,Puri1994}
\begin{equation}
\Delta \mathcal{O}^2<|\langle C\rangle|/2.\label{definition1}
\end{equation}
Following this line of thought, we can generalize this definition to the case of the variances satisfying the stronger inequality
(\ref{inequality3}), that is, one can define squeezing if
\begin{equation}
\Delta \mathcal{O}^2<|\langle C\rangle|/(2-\Theta_1).\label{definition2}
\end{equation}
Apparently, the definition (\ref{definition2}) will reduce to that of (\ref{definition1}) if $\Theta_1=0$, which
means $|\Psi\rangle$ belongs to the OISs. Alternatively, by use of the inequality (\ref{inequality1}), another
criterion of squeezing can be given as
\begin{equation}
\Delta \mathcal{O}^2<\frac{|\langle C\rangle|+\Theta_2}{2}.
\end{equation}
As expected, when $\Theta_2=0$ this definition also reduce to the one based on the RUR.
Furthermore, some remarks are in order: (i) the definition of squeezing is not unique.
An appropriate criterion should be established depending on the specific scenario
where this definition makes sense \cite{Walls1981}; (ii) given $A$ and $B$, the values
of $\Theta_1$ and $\Theta_2$ are determined by the choice of $|\Psi^{\perp}\rangle$ \cite{Maccone2014}.
More precisely, the possible choices of $|\Psi^{\perp}\rangle$ will decide the strength of the definition.

In particular, in the context of spin squeezing, various spin squeezing parameters have been proposed
for different applications, which attracted increasing attention since spin squeezing has been recognized
as a valuable resource for quantum metrology \cite{Wineland1992,Kitagawa1993,Pezze2009,Hyllus2010}.
For instance, based on the definition (\ref{definition1}), a spin squeezing parameter can be defined
with respect to two orthogonal unit vectors $\vec{n}_1$ and $\vec{n}_2$ \cite{Ma2011}
\begin{equation}
\xi_H^2=2(\Delta \bm{J}_{n_1})^2/|\langle \bm{J}_{n_2}\rangle|,
\end{equation}
where $\bm{J}_{n}=\bm{J}\cdot\vec{n}$, angular momentum operator $\bm{J}=\frac{1}{2}\sum_{l=1}^{N}\vec{\sigma}^{(l)}$
and $\vec{\sigma}^{(l)}$ is the vector of Pauli matrices acting on the $l$th particle. When $\xi_H^2<1$, the state
is said to be squeezed. However, $\xi_H^2$ may be less than $1$ in coherent spin state (CSS) and this is not desirable
since a CSS should not be viewed as being spin-squeezed \cite{Wineland1992,Ma2011,Arecchi1972}.
Therefore, we shift our focus to the squeezing parameter introduced by Wineland \textit{et al.},
which is the one directly related to interferometric sensitivity \cite{Wineland1992}
\begin{equation}
\xi_R^2=N(\Delta \bm{J}_{n_1})^2/|\langle \bm{J}_{n_2}\rangle|^2.
\end{equation}
For the scenario of optical interferometry, it has been proved that \cite{Pezze2009,Hyllus2010}
\begin{equation}
\xi_R^2=\frac{N(\Delta \bm{J}_{n_1})^2}{|\langle \bm{J}_{n_2}\rangle|^2}\geq
\frac{N}{\mathcal{F}[|\Psi\rangle,\bm{J}_{n_3}]}=\chi^2,\label{ancilla6}
\end{equation}
where $\mathcal{F}[|\Psi\rangle,\bm{J}_{n_3}]=4(\Delta \bm{J}_{n_3})^2$ is the quantum Fisher information \cite{Braunstein1994}
and $\vec{n}_3$ is orthogonal to both $\vec{n}_1$ and $\vec{n}_2$. The parameter $\chi^2<1$ is a sufficient condition for
particle entanglement and Eq. (\ref{ancilla6}) confirms that there exists a class of states which are entangled, $\chi^2<1$,
but not spin squeezed \cite{Strobel2014}. In fact, by applying uncertainty relation (\ref{inequality3}), we can obtain
a generalized version of Eq. (\ref{ancilla6})
\begin{equation}
\xi_R^2\geq\frac{\chi^2}{(1-\Theta_1/2)^2}\geq\chi^2,
\end{equation}
where we choose $A=\bm{J}_{n_1}$, $B=\bm{J}_{n_3}$ and $[\bm{J}_{n_3},\bm{J}_{n_1}]=i\bm{J}_{n_2}$.
Hence, if we want to attain higher sensitivity (e.g., $\xi_R^2$ as small as possible), we are supposed to
choose the input state within the set of OISs ($\Theta_1=0$) \cite{Hillery1993}.

\section{CONCLUSIONS}\label{sec5}
In this work, we construct and formulate two quantum uncertainty \textit{equalities}, which hold for all pairs of incompatible observables
and lead to the new uncertainty relations recently introduced by Maccone and Pati \cite{Maccone2014}.
In fact, one can obtain a series of inequalities \textit{with hierarchical structure} by retaining $1$ to $d-2$ terms
within the set $\{|\Psi^\perp_{k}\rangle_{k=1}^{d-1}\}$. Remarkably,
we provide an explicit interpretation lying behind the structure of these inequalities and relate them to the so-called intelligent states
\cite{Aragone1974,Aragone1976}. As an illustration, we investigate the properties of these uncertainty inequalities in the qubit system and
a state-independent bound is obtained for the sum of variances. Finally, the implication of these uncertainty relations
in interferometric sensitivity is also discussed in the context of spin squeezing.

Possible generalizations of our method need to be addressed. First, here we only consider the extension of RUR, but one can also
extend the SUR employing the concept of GISs \cite{Trifonov1994}, where a suitable phase factor $e^{i\omega}$ should be introduced.
For example, we can establish a strengthened version of the inequality (\ref{inequality1})
\begin{equation}
\Delta A^2+\Delta B^2\geq\sqrt{|\langle C\rangle|^2+|\langle F\rangle|^2}+\Theta_3.
\end{equation}
where
\begin{align}
\Theta_3=\left|\langle\Psi|A\pm ie^{i\omega}B|\Psi^\perp\rangle\right|^2,\\
\gamma=e^{i\omega},\quad \tan\omega=\langle F\rangle/\langle C\rangle.
\end{align}
Moreover, we notice that Pati and Wu extended the result of \cite{Maccone2014} into the realm of weak measurement \cite{Pati2014}.
In fact, the corresponding uncertainty \textit{equality} of Eq. (6) in \cite{Pati2014} can also be constructed utilizing
our approach.

\begin{acknowledgments}
Y. Y. is very grateful to Dr. L. Ge for many helpful discussions.
This research is supported by the National Natural Science Foundation of China (Grants No. 11025527, No. 11121403, No. 10935010, No. 11074261, and No. 11247006),
the National 973 program (Grants No. 2012CB921602, No. 2012CB922104, and No. 2014CB921403), and the China Postdoctoral Science Foundation (Grant No. 2014M550598).
\end{acknowledgments}
\appendix

\section{OIS, GIS and minimum-uncertainty states}\label{appendix1}
As mentioned above, the OISs and GISs are quantum states that satisfy the equality sign in
the RUR and SUR, respectively. However, frequently the OISs and GISs are also termed as
\textit{minimum-uncertainty states} in previous literature \cite{Trifonov2000}.
Obviously, there is no commonly accepted name for those states and here we prefer to call
the states that minimize the product functional $\mathcal{U}(\Psi)=\Delta A^2\Delta B^2$
as minimum-uncertainty states \cite{Wodkiewicz1985}. In Ref. \cite{Jackiw1968}, Jackiw
also termed it as \textit{critical states} and presented a necessary condition which
must be satisfied if $\mathcal{U}(\Psi)$ achieves the minimum value
\begin{equation}
\left(\frac{\breve{A}^2}{\Delta A^2}+\frac{\breve{B}^2}{\Delta B^2}-2\right)|\Psi\rangle=0,\label{app1}
\end{equation}
where $\breve{\mathcal{O}}=\mathcal{O}-\langle\mathcal{O}\rangle I$.

In fact, we can also provide the similar constraint for the sum functional $\mathcal{W}(\Psi)=\Delta A^2+\Delta B^2$
by the variational method. Considering the variation $\langle\Psi|\rightarrow\langle\Psi|+\langle\delta\Psi|$,
we have
\begin{align}
\delta(\langle\mathcal{O}\rangle)\approx\frac{\langle\delta\Psi|\mathcal{O}|\Psi\rangle}{\langle\Psi|\Psi\rangle}
-\frac{\langle\Psi|\mathcal{O}|\Psi\rangle}{\langle\Psi|\Psi\rangle^2}\langle\delta\Psi|\Psi\rangle,
\end{align}
where only the first-order approximation is adopted.
Therefore, the variation of the variance can be represented as
\begin{align}
\delta(\Delta\mathcal{O}^2)&=\delta(\langle\mathcal{O}^2\rangle-\langle\mathcal{O}\rangle^2)\nonumber\\
&=\delta(\langle\mathcal{O}^2\rangle)-2\langle\mathcal{O}\rangle\delta(\langle\mathcal{O}\rangle)\nonumber\\
&\approx\langle\delta\Psi|\left[\frac{(\mathcal{O}-\langle\mathcal{O}\rangle)^2|\Psi\rangle}{\langle\Psi|\Psi\rangle}\right]
-\Delta\mathcal{O}^2\frac{\langle\delta\Psi|\Psi\rangle}{\langle\Psi|\Psi\rangle}\nonumber\\
&=\frac{\langle\delta\Psi|\breve{\mathcal{O}}^2|\Psi\rangle}{\langle\Psi|\Psi\rangle}
-\Delta\mathcal{O}^2\frac{\langle\delta\Psi|\Psi\rangle}{\langle\Psi|\Psi\rangle}¡£
\end{align}
Note that the normalization factor $\langle\Psi|\Psi\rangle$ plays an important role in the derivation.
By choosing $\mathcal{O}=A,B$, the stationary condition $\delta[\mathcal{W}(\Psi)]=0$ can be recast as
\begin{equation}
\langle\delta\Psi|(\breve{A}^2+\breve{B}^2)|\Psi\rangle=\langle\delta\Psi|(\Delta A^2+\Delta B^2)|\Psi\rangle.
\end{equation}
For the arbitrariness of $\langle\delta\Psi|$, it follows that
\begin{equation}
(\breve{A}^2+\breve{B}^2)|\Psi\rangle=(\Delta A^2+\Delta B^2)|\Psi\rangle.
\end{equation}
This condition defines another class of \textit{critical states} for the functional $\mathcal{W}(\Psi)$.
Furthermore, combining with the extra constraint $\Delta A=\Delta B$, this condition reduces to
a special case of Eq. (\ref{app1}), which is to be expected.

\section{State-independent uncertainty relation for qubit system}\label{appendix2}
In this Appendix, we aim to prove the following \textit{state-independent} bound
\begin{equation}
\mathcal{W}(\Psi)\geq1-|\vec{a}\cdot\vec{b}|=2(1-c^2).
\end{equation}
Indeed, this relation holds for arbitrary (pure or mixed) single-qubit states.
Before proceeding, we would like to present two facts to simplify our discussion:
(i) $\mathcal{W}(\Psi)$ reaches the minimum value for pure states; (ii) we only
need to consider the states whose Bloch vector $\vec{r}$ lies within the plane
spanned by $\vec{a}$ and $\vec{b}$. The point (i) is obvious since we have
$\mathcal{W}(\Psi)=2-(\vec{a}\cdot\vec{r})^2-(\vec{b}\cdot\vec{r})^2$.
Thus, we have
\begin{equation}
\min_{|\vec{r}|\leq1}\mathcal{W}(\Psi)=\min_{|\vec{r}|=1}\mathcal{W}(\Psi).
\end{equation}
In addition, if $\vec{r}$ is not coplanar with $\vec{a}$ and $\vec{b}$, we have
the following relation
\begin{equation}
\cos\vartheta=\cos\vartheta_1\cos\vartheta_2.\label{app2}
\end{equation}
where $\vec{r}\vee\vec{r}_\parallel=\vartheta_1$, $\vec{r}_\parallel\vee\vec{a}=\vartheta_1$
and $\vec{a}\vee\vec{r}=\vartheta$. Here $\vec{m}\vee\vec{n}$ denotes the angle between
the two vectors and $\vec{r}_\parallel$ represents the unit vector along the projection of $\vec{r}$
on the plane spanned by $\vec{a}$ and $\vec{b}$. From Eq. (\ref{app2}), we obtain $|\cos\vartheta_1|\geq|\cos\vartheta|$,
that is, $(\vec{a}\cdot\vec{r}_\parallel)^2\geq(\vec{a}\cdot\vec{r})^2$. Note that the same argument applies to $\vec{b}$.
Therefore, the minimum value of $\mathcal{W}(\Psi)$ is attained within this plane.

Since $|\vec{r}|=1$, $\mathcal{W}(\Psi)$ can be written as
\begin{equation}
\mathcal{W}(\Psi)=\|\vec{a}\times\vec{r}\|^2+\|\vec{b}\times\vec{r}\|^2.
\end{equation}
Moreover, the Bloch vector $\vec{r}$ can be decomposed as
\begin{equation}
\vec{r}=\alpha\frac{\vec{a}+\vec{b}}{\|\vec{a}+\vec{b}\|}+\beta\frac{\vec{a}-\vec{b}}{\|\vec{a}-\vec{b}\|},
\end{equation}
where $\alpha,\beta$ are real parameters and $\alpha^2+\beta^2=1$.
By using the parallelogram law in Herbert space, we have
\begin{align}
\mathcal{W}(\Psi)&=\|\vec{a}\times\vec{r}\|^2+\|\vec{b}\times\vec{r}\|^2 \nonumber\\
&=(\|(\vec{a}+\vec{b})\times\vec{r}\|^2+\|(\vec{a}-\vec{b})\times\vec{r}\|^2)/2 \nonumber\\
&=[\beta^2(2+2\vec{a}\cdot\vec{b})+\alpha^2(2-2\vec{a}\cdot\vec{b})]/2 \nonumber\\
&=1+\vec{a}\cdot\vec{b}(\beta^2-\alpha^2).
\end{align}
Since $|\beta^2-\alpha^2|\leq1$, we finally obtain $\mathcal{W}(\Psi)\geq1-|\vec{a}\cdot\vec{b}|$.
This inequality can be rewritten as
\begin{equation}
(\vec{a}\cdot\vec{r})^2+(\vec{b}\cdot\vec{r})^2\leq2c^2=1+|\vec{a}\cdot\vec{b}|,
\end{equation}
where $c=\max_{i,j}|\langle a_i|b_j\rangle|$ and $\{|a_i\rangle\}$ ($\{|b_j\rangle\}$) are the corresponding eigenvectors of $A$ ($B$).
We notice the similar bounds have been obtained in \cite{Busch2014b} and \cite{Mandayam2014}. However, Ref. \cite{Busch2014b}
does not provide an explicit proof and comparing with \cite{Mandayam2014}, our proof is compact and straightforward.


\end{document}